\documentclass[fleqn,usenatbib]{mnras}

\usepackage{newtxtext,newtxmath}
\usepackage[T1]{fontenc}
\usepackage{ae,aecompl}

\usepackage{graphicx}	% Including figure files
\usepackage{amsmath}	% Advanced maths commands

\usepackage{amssymb}	% Extra maths symbols
\usepackage[normalem]{ulem}

\title[Removing Interlopers from LIM Non-Gaussianity]{Removing Interlopers From Intensity Mapping Probes Of  Primordial Non-Gaussianity}

\author[Chang Chen et al.]{
Chang Chen,$^{1}$\thanks{E-mail: changchen@nyu.edu}
Anthony R. Pullen,$^{1,2}$
\\
$^{1}$Center for Cosmology and Particle Physics, Department of Physics, New York University, 726 Broadway, New York, NY, 10003, U.S.A.\\
$^{2}$Center for Computational Astrophysics, Flatiron institute, New York, NY 10010, U.S.A.
}

\date{Accepted XXX. Received YYY; in original form ZZZ}

\pubyear{2015}

\begin{document}
\label{firstpage}
\pagerange{\pageref{firstpage}--\pageref{lastpage}}
\maketitle

% Abstract of the paper
\begin{abstract}
Line intensity mapping (LIM) has the potential to produce highly precise measurements of scale-dependence bias from primordial non-Gaussianity (PNG) due to its ability to map much larger volumes than are available from galaxy surveys.   PNG parameterized by $f_{NL}$ leads to a scale-dependent correction to the bias, and therefore a correction to the line intensity power spectrum.  However, LIM experiences contamination from foreground emission, including interloping emission lines from other redshifts which alter the power spectra of the maps at these scales, potentially biasing measurements of $f_{NL}$.  Here we model the effect of line interlopers on upcoming line intensity mapping probes of primordial non-Gaussianity (PNG) from inflation. As an example, we consider the $[\rm CII]$ line at target redshift $z_t = 3.6$ to probe PNG, with the important systematic concern being foreground contamination from CO lines residing at redshifts different from the target redshift. We find interloper lines can lead to a significant bias and an increase in errors for our PNG constraints, leading to a false positive for non-standard inflation models. We model how well the cross-correlation technique could reduce this interloper contamination. We find the uncertainty of $f_{NL}$ reduces by factors of two and six for local and orthogonal shape PNG respectively, and by a factor of five for local shape if we consider seven interloper lines, almost eliminating the effect of interlopers. This work shows that using cross-power and auto-power spectra of line intensity maps jointly could potentially remove the effects of interlopers when measuring non-Gaussianity.
\end{abstract}

\begin{keywords}
large-scale structure of Universe --  diffuse radiation -- inflation
\end{keywords}

%%%%%%%%%%%%%%%%%%%%%%%%%%%%%%%%%%%%%%%%%%%%%%%%%%

%%%%%%%%%%%%%%%%% BODY OF PAPER %%%%%%%%%%%%%%%%%%

\section{INTRODUCTION}

Inflation theory has been enormously successful in explaining the origin of large-scale structure in the universe \citep{1981PhRvD..23..347G, 1987quco.book..149L}. The simplest inflation model predicts almost scale-invariant, adiabatic and Gaussian perturbations. Primordial Non-Gaussianity (PNG) of the local shape is parameterized by $f_{NL}^{loc}$ \citep{PhysRevD.63.063002, 1994ApJ...430..447G} as
\begin{equation}
\Phi = \phi + f_{NL}^{loc}\phi^2
\end{equation}
where $\phi$ is the Gaussian primordial potential. The parameter $f_{NL}$ for the standard single-field inflation model is very small, of order $10^{-2}$ \citep{2003JHEP...05..013M}, while $f_{NL}^{loc}$ for more general models, e.g. multi-field inflation\citep{1997PhRvD..56..535L, 2003PhRvD..67b3503L} is much higher, even up to $f_{NL}\sim\mathcal{O}(1)$. In addition, the orthogonal shape PNG, which can be produced by models with higher derivative interactions and Galilean inflation \citep{2010JCAP...01..028S}, is parameterized by $f_{NL}^{orth}$. Therefore, probing $f_{NL}$ can serve as a strong discriminant among cosmological models.\par

Currently, the best constraints on $f_{NL}$ are from CMB bispectra measurements from Planck satellite, given as $f_{NL}^{loc}= -0.9 \pm 5.1$ and $f_{NL}^{orth}= -38 \pm 24$ \citep{2019arXiv190505697P}. Complimentary large-scale structure (LSS) surveys such as BOSS \citep{2017MNRAS.470.2617A}, DES \citep{2018PhRvD..98d3526A}, EUCLID \citep{2018LRR....21....2A} could probe $f_{NL}$ scale-dependent bias on large scales \citep{2008ApJ...677L..77M, 2008PhRvD..78l3507A} and are promising to achieve higher precision than CMB data.As of now, current constraints on $f_{NL}^{loc}$ from scale-dependent bias are not yet competitive with CMB constraints \citep{2014PhRvL.113v1301L,2019JCAP...09..010C,2020MNRAS.491L..61D}, with the most precise constraints still coming from quasars in \citet{2014PhRvL.113v1301L}, $-49 <f_{NL}^{loc}<31$. For PNG of the general shape, non-zero primordial bispectrum induces a scale-dependent correction to the bias \citep{2008PhRvD..77l3514D, 2008ApJ...677L..77M, 2008PhRvD..78l3507A,2011PhRvD..84f3512D}, leaving a signature in the power spectrum on large scales.\par
Unlike galaxy surveys which probe individual tracers, line intensity mapping measures cumulative intensity fluctuations from all luminous sources including faint ones unresolved in galaxy surveys. Line intensity mapping experiments could have the potential to access over $80\%$ of the volume of the observable Universe \citep{2019BAAS...51c.101K}. Therefore, it has the potential to further the study of large-scale structure in the universe, especially of the Epoch of
Reionization when sources with low luminosity inaccessible
to galaxy surveys reionized the universe \citep{2015ApJ...802L..19R}. The fine-structure line from ionized
carbon ([CII]) \citep{2012ApJ...745...49G, 2015ApJ...806..209S}, the brightest line for star formation galaxies, makes it an ideal candidate for probing the line intensity power spectrum on large scales to improve constraints of PNG \citep{2019ApJ...872..126M, 2019ApJ...870L...4M}. \par

An important systematic effect when considering intensity mapping is foreground interloper emission \citep{2010JCAP...11..016V, 2016ApJ...825..143L} from lines residing at the same \emph{observed} frequency as the target line. Interloper emission, in which different emission lines from galaxies with different epochs redshift to the same observed frequency, contaminates intensity maps and introduces extra correlations in the measured power spectrum that can bias parameter estimations.  This deposits a distinct signature at large scales where effects from non-Gaussianity reside due to the features in the power spectrum at these scales such as baryon acoustic oscillations and the peak corresponding to matter-radiation equality.

In our case, we consider a target line of $[\rm CII]$ emission with rest-frame frequency $\nu_{t} = 1901$ GHz emitted from sources at redshift $z_{t}$ contaminated by interloping rotational lines of carbon monoxide ($\rm CO\left( J \rightarrow J-1 \right)$) \citep{2010JCAP...11..016V} with rest-frame frequency $\nu_{i} = 115 \times J GHz$ emitted from sources at redshifts $z_{i}$ lower than target redshift $z_{t}$. Those CO lines can be confused as $[\rm CII]$ line if they reside at the same observed frequency $\nu_{obs}$ as
the target line, i.e. $\nu_{obs}=\nu_{t}/(1+z_t)=\nu_{i}/(1+z_i)$. These interloper CO lines can distort power spectrum detections and contaminate the forecast for constraints on PNG. A recent paper \citep{2020ApJ...894..152G} discusses the interloper contamination on constraints of cosmological and astrophysical parameters. Assuming the target redshift $z_{t}$ in converting the
observed frequencies and angles to their co-moving coordinates,
the interloper lines will be mapped to the
wrong longitudinal and transverse co-moving wavenumbers, causing an anisotropy of the
interloper power spectrum
contribution \citep{2010JCAP...11..016V,   2014ApJ...785...72G, 2016ApJ...825..143L}.

In this paper we adopt this approach, where we seek how best to reduce the contamination when measuring $f_{\rm NL}$ using power spectrum measurements. In this work we consider a line intensity mapping survey using a Planck-like telescope with a high-resolution spectrograph, similar to the instrument modeled in \citet{2019ApJ...870L...4M}, to forecast measurement errors for $f_{\rm NL}$. We find through a Fisher analysis that not only do interlopers significantly increase the error on $f_{\rm NL}$, but that a properly modeled auto-power spectrum analysis cannot reduce the error effectively. Specifically, we find that interlopers can bias the $f_{NL}$ measurement, producing a false 3$\sigma$ detection of non-Gaussianity favoring non-standard inflation models. \par

Next, we consider cross-correlation to separate anisotropic interloper emission from the $[\rm CII]$ auto power spectrum.  We model the cross-correlation between $[\rm CII]$ and CO(4-3) at the same redshift, $z=3.6$ in our case with a different frequency
to avoid bias from interloper emission. We calculate auto-power spectra and cross-power spectrum of $[\rm CII]$ and CO(4-3) maps and perform a Fisher analysis to forecast the errors. We find the interloper contamination for the PNG probing can be largely removed using this method, reducing the Fisher analysis error by factors of two and five for $f_{NL}^{loc}$ and $f_{NL}^{orth}$ respectively if we only consider 1 interloper line, and by a factor of five for $f_{NL}^{loc}$ if we consider 7 interloper lines.  For both cases we find that including the cross-power in the analysis removes most of this interloper bias, making the $f_{\rm NL}$ measurement rather insensitive to contamination from interlopers. However, we also find that the auto-power spectra can still play a large role in regards to the precision of the $f_{\rm NL}$ measurement.  Specifically, we find that for $f_{\rm NL}^{loc}$, the precision comes solely from the cross-power spectrum, while for $f_{\rm NL}^{ortho}$ the precision comes roughly equally from the auto-power and cross-power spectra.\par

The outline of the rest of the paper is as follows. In Section \ref{S:PNGLIM}, we review the theoretical model of the line intensity power spectrum detection for PNG, and the interloper contamination for this detection. In Section \ref{S:PEF}, we describe the Fisher analysis and data bias approach. In Section \ref{S:FSR}, we detail our fiducial survey and present the results for constraints on $f_{NL}^{loc}$. We further discuss the prospect for using cross-correlation with another emission line in Section \ref{S:CROSS}. In Section \ref{S:Orth} we 
forecast the constraints on the orthogonal shape PNG. We further consider different astrophysical models in Section \ref{S:DISCUSSION}. Based on our results, we draw our conclusions in Section \ref{S:CONCLUSION}.  Throughout this paper we adopt a flat cosmological model with $H_0 = 67.74, \Omega_m = 0.26,\Omega_b = 0.049, \sigma_8 = 0.816$.

\section{PNG LINE INTENSITY POWER SPECTRUM  DETECTION AND  INTERLOPER CONTAMINATION} \label{S:PNGLIM}
In this section we review the formalism for the scale-dependent correction to clustering bias due to non-zero $f_{NL}$, present how this correction to bias would manifest itself in the line intensity power spectrum detection, and highlight the interloper contamination of the power spectrum.

\subsection{PNG review}
Local shape PNG is produced by super-horizon non-linear evolution of primordial curvature perturbations.
In the absence of PNG, on large scales the halo bias $b_h$ is assumed to be constant. Local shape PNG, parameterized by $f_{NL}^{\rm loc}$, leads to a scale-dependent correction to the linear halo bias \citep{2008PhRvD..77l3514D, 2008ApJ...677L..77M, 2008PhRvD..78l3507A}.
\begin{equation}
b_h(M,z) \rightarrow b_h(M,z) + \Delta b_{h}^{\rm loc}(M,k,z)
 \label{eq:biascorrection1}
\end{equation}
\begin{equation}
\Delta b_{h}^{\rm loc}(M,k,z) =   \frac{3f_{\rm NL}^{\rm loc}  \delta_c \left[b_{h}(M,z)-1\right]\Omega_m H_0^2}{k^2T(k)D(z)}
 \label{eq:biascorrection2}
\end{equation}
where $\delta_c = 1.686$ is the threshold of spherical collapse at $z = 0$, $T(k)$ is the matter linear transfer function $T(k\rightarrow0) = 1$ and $D(z)$ is the normalized linear growth factor [$D(z=0) = 1$]. 

On large scales this $k^{-2}$ scale-dependent correction has been used to constrain  $f_{\rm NL}^{\rm loc}$ from power spectrum detections of LSS biased tracers due to its clean signal \citep{2008JCAP...08..031S,2014PhRvL.113v1301L,2015JCAP...05..040H}. Alternatively, line intensity mapping can provide a more economical survey over larger volume to probe power spectrum on large scales.\par

Orthogonal shape PNG can be produced by models with higher-derivative interactions and Galilean inflation \citep{2010JCAP...01..028S}. Parameterized by $f_{NL}^{orth}$, on large scales, orthogonal shape PNG leads to a scale-dependent correction to the linear halo bias \citep{2011PhRvD..84f3512D}
\begin{eqnarray}
\label{eq:biascorrection3}
\Delta b_h^{orth}(k) &=&   -6f_{NL} \frac{\sigma_{\alpha s}^2}{\sigma_{0 s}^2}\left[(b_h-1)\delta_c + 2\left(\frac{\partial ln \sigma_{\alpha s}}{\partial ln \sigma_{0 s}}-1\right)\right]\nonumber\\
&&\times k^{-2\alpha}\mathcal M_{R_s}(k,z)^{-1}
\end{eqnarray}
$\alpha=(n_s-4)/6$, $\mathcal M_{R_s}(k,z) = W_{R_s}(k)\mathcal M(k,z)$, where $\mathcal M(k,z)= \frac{2}{5} \frac{k^2T(k)D(z)}{\Omega_m H_0^2}$, and we choose the window function $W_{R_s}(k)$ as the Fourier transform of a spherical top-hat filter with smoothing length $R_s=(3M/4\pi\bar\rho)^{1/3}$,
\begin{equation}
W_{R_s}(k) = \frac{3\left[{sin}(kR_s) - kR \  {cos} (kR_s)\right]}{(kR_s)^3}
\end{equation}
the general spectrum moment is \begin{equation}
\sigma^2_{\alpha s}=\frac{1}{2\pi^2}\int^{\infty}_0 dk k^{2(\alpha +1)}P_{\Phi}(k) \mathcal M_{R_s}(k,z)^{-1}
\end{equation}

\subsection{Line intensity power spectrum}
Here we model the $[\rm CII]$ line intensity power spectrum to predict how well it could constrain PNG. The relation between the mean intensity of the emission line and
the luminosity of $[\rm CII]$-luminous galaxies in their host halos \citep{2010JCAP...11..016V, 2019ApJ...872..126M} at redshift z is expressed as
\begin{equation}\langle I_{\rm CII}\rangle (z)  = \frac{c^2 f_{\rm duty}}{2k_B \nu_{\rm obs}^2} \int_{M_{\rm min}}^{M_{\rm max}} dM \frac{dn}{dM} \frac{L(M,z)}{4 \pi \mathcal{D}_{L}^{2}} \left ( \frac{dl}{d\theta} \right )^{2} \frac{dl}{d\nu}
\label{eq:intensity}
\end{equation} 
where $M_{\rm min}$ and $M_{\rm max}$ are the minimum and maximum mass of $[\rm CII]$-emitting halos, $f_{\rm duty}$ is the duty cycle of line emitting halos, (i.e. the fraction of halos in the given mass range which emit $[\rm CII]$ line at redshift z), $dn/dM$ is the halo mass function, $L(M,z)$ is the luminosity of the $[\rm CII]$-luminous galaxies from dark matter halo of mass M at redshift $z$, ${\mathcal D}_L$ is the luminosity distance. We set $M_{\rm min} = 10^{9}M_\odot$ and $M_{\rm max} = 10^{14.5}M_\odot$. We replace $f_{\rm duty}$ with $p_{n,\sigma}$ as given in \citep{2019ApJ...872..126M}. We use the Tinker halo mass function \citep{2008ApJ...688..709T}. $dl/d\theta$ and $dl/d\nu$ convert comoving lengths $l$ to frequency $\nu$ and angular size $\theta$, $dl/d\theta = D_{\rm A, co}(z)$, $\frac{dl}{d\nu} = \frac{c(1+z)}{\nu_{\rm obs}H(z)}$, where $D_{\rm A, co}(z)$ is the comoving angular diameter distance (For a flat universe, $D_{\rm A,co}(z) = \chi(z)$ where $\chi(z)$ is the co-moving distance to redshift $z$.) and $H(z)$ is the Hubble parameter at redshift z. For our fiducial model we use the result of the $m1$ model of \citet{2015ApJ...806..209S} to relate the $[\rm CII]$ luminosity to the average star formation rate $\overline{\rm SFR}(M,z)$ of \citet{2013ApJ...762L..31B}
\begin{equation}
\log L_{\rm CII} = 0.8475 \times \log \overline{\rm SFR}(M,z) + 7.2203
\end{equation}
$[\rm CII]$ line bias is related to halo bias $b_h(M,z)$ as \citep{2019ApJ...872..126M}
\begin{equation}
b_{\rm line}(z) = \frac{\int_{M_{\rm min}}^{M_{\rm max}} dM  \ \frac{dn}{dM} \  b_h(M,z) L(M,z)  
}{\int_{M_{\rm min}}^{M_{\rm max}} dM \ \frac{dn}{dM} \ L(M,z)}
\label{eq:bias}
\end{equation}
We use halo bias $b_h(M,z)$ of \citep{2010ApJ...724..878T}.\par
Setting $z = 3.60$, we get $\langle I_{\rm CII}(3.6) \rangle= 0.30\mu K$ from~(\ref{eq:intensity}) and $b_{\rm line}(3.6) = 3.47$ from~(\ref{eq:bias}).\par
The shot noise power spectrum is
\begin{equation}
\label{eq:pshot}
P_{\rm shot}(z) = \frac{c^4 f_{\rm duty}}{4k_B^2 \nu_{\rm obs}^4}  \int_{M_{\rm min}}^{M_{\rm max}} dM \frac{dn}{dM} {\left[\frac{L(M,z)}{4 \pi \mathcal D_L^2} 
\left ( \frac{dl}{d\theta} \right )^{2} \frac{dl}{d\nu} \right ]}^2.
\end{equation}
Including redshift space distortions, the line clustering power spectrum can be expressed as 
\begin{align}
P(k,\mu,z) &= \langle I_{\rm CII}\rangle (z)^2 b_{\rm line}^2(z) P_0(k,z)\nonumber \\
&\times \left[1+\mu^2\beta(k,z)\right]^2{\rm exp}\left(-\frac{k^2 \mu^2 \sigma_v^2}{H^2(z)}\right)
\label{eq:power}
\end{align}
The factor of $\left[1+\mu^2\beta(k,z)\right]^2$ comes from the Kaiser effect \citep{1987MNRAS.227....1K}, $\beta(k,z) = f/b_{\rm line}(z)$,where $f$ is the logarithmic derivative of the growth factor, $\mu = k_{\parallel} / k$ is the cosine of the angle between the k and the line of sight direction. The factor of ${\rm exp}\left(-\frac{k^2 \mu^2 \sigma_v^2}{H^2(z)}\right)$ is from the finger-of-god effect \citep{1972MNRAS.156P...1J} with the pairwise velocity dispersion approximated as $\sigma_v^2 = (1+z)^2\left[\frac{\sigma_{\rm FOG}^2(z)}{2} + c^2 \sigma_z^2\right]$, where $\sigma_{\rm FOG}(z) = \sigma_{{\rm FOG},0}\sqrt{1+z}$, $\sigma_z = 0.001 (1+z)$ and $\sigma_{{\rm FOG},0}=250 km.s^{-1}$.
According to Eqs.~(\ref{eq:bias}),~(\ref{eq:biascorrection1}) and~(\ref{eq:biascorrection2}) will manifest as a scale-dependent correction to the $b_{\rm line}$ and therefore to the line power spectrum~(\ref{eq:power}).
\subsection{Interloper contamination}
Line emission of CO interloper line at redshift $z_i$ confuses with target $[\rm CII]$ line by residing at the same observed frequency $\nu_{obs}$ as the $[\rm CII]$ line. When we convert the observed frequency interval $\Delta \nu_{\rm obs}$ and angular separation $\Delta \theta_{obs}$ in a data cube to co-moving coordinates to calculate the power spectrum, if we adopt the target line redshift $z_t$, we will end up with wrong wavenumbers \citep{2016ApJ...825..143L}. The relation between the true wavenumbers $k_\parallel$ and $\textbf{k}_\perp$ and these apparent wavenumbers $\tilde{k}_\parallel$ and $\tilde{\textbf{k}}_\perp$, calculated by wrongly adopting the $z_t$ for coordinate conversion, is
\begin{equation}
\tilde{k}_\parallel = \frac{H(z_t)}{H(z_i)} \frac{1+z_i}{1+z_t} k_\parallel = \alpha_\parallel k_\parallel
\end{equation}
\begin{equation}
\tilde{\textbf{k}}_\perp = \frac{D_{\rm A, co}(z_i)}{D_{\rm A, co}(z_t)} \textbf{k}_\perp = \alpha_\perp \textbf{k}_\perp
\end{equation}
where $\alpha_\parallel$ and $\alpha_\perp$ are distortion factors caused by this incorrect redshift adoption.
Therefore for the multiple interloper line emission case with N interloper lines, the total power spectrum in the data cube is
\begin{align}
\label{eq:ptot}
P_{\rm tot}(\tilde{k}_\parallel,\tilde{\textbf{k}}_\perp) =& P_t(k_\parallel,\textbf{k}_\perp) \nonumber \\
& + \sum_{j=1}^{N} \frac{1}{\alpha_\parallel(z_j) \alpha_\perp^2(z_j)} P_{j} \left(\frac{\tilde{k}_\parallel}{\alpha_\parallel},\frac{\tilde{\textbf{k}}_\perp}{\alpha_\perp}\right).
\end{align}
where $z_j$ is the redshift of the $j$th interloper line.
In this paper we consider up to 7 CO interloper lines, with $J$ ranging from 4 to 10 in $\rm CO\left( J \rightarrow J-1 \right)$ molecule transitioning between rotational states $J$ and $J-1$.\par
For $\rm CO\left( 1 \rightarrow 0 \right)$, expressed in units of solar luminosity, the luminosity is 
\begin{equation}
L_{\rm CO(1-0)} = 4.9 \times 10^{-5}  L'_{\rm CO}
\end{equation}
Using the fit from \citep{2013ARA&A..51..105C}, CO line luminosity $L'_{\rm CO}$, which is expressed in units of $\rm K  kms^{-1}pc^2$, is related to the far-infrared luminosity $L_{IR}$, which is in the unit of $L_\odot$ as 
\begin{equation}
\log L_{\rm IR} = 1.37 \times \log  L'_{\rm CO} -  1.74
\end{equation}
We use the base model of \citet{2016ApJ...817..169L}, where $L_{\rm IR}$ is related to the $\overline {\rm SFR}$ in units of $M_\odot {\rm yr}^{-1}$ via Kennicut relation \citep{1998ARA&A..36..189K} using the results of \citet{2013ApJ...762L..31B}
\begin{equation}
{\overline {\rm SFR}}(M,z) = \delta_{\rm MF} \times 10^{-10} L_{\rm IR}.
\end{equation}
We use $\delta_{\rm MF} = 1$ \citep{2013ApJ...762L..31B, 2016ApJ...817..169L}.\par
Using Eq.~(\ref{eq:intensity}) we get intensities for $\rm CO\left( 1 \rightarrow 0 \right)$ at different redshifts $z_j$  that will interlope with the $[\rm CII]$ target line at $z = 3.6$, i.e. $\frac{\rm J \times 115 \rm GHz}{z_j + 1} = \frac{1901 \rm GHz}{3.6 + 1}$. In order to construct a rough estimate of the interloper luminosities, we use the ratio between the luminosity of 7 CO interloper lines with $J$ ranging from 4 to 10  and that for the $\rm CO\left( 1 \rightarrow 0 \right)$  \citep{2010JCAP...11..016V} to get the intensities for these 7 interlopers, we also calculate the line biases for these interloper lines according to Eq.~(\ref{eq:bias}); these values for interlopers are listed in Table~\ref{tab:values}.\par
\begin{figure}
	\includegraphics[width=\columnwidth]{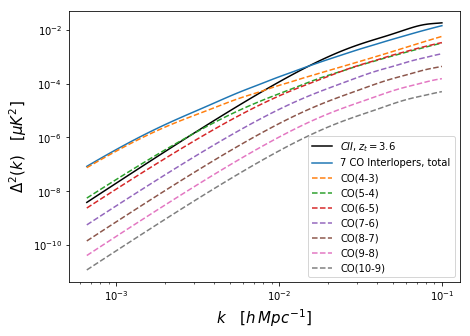}
    \caption{Spherically averaged $k^3 P(k)/(2\pi)^2$. The black solid line shows the $[\rm CII]$ line power spectrum at $z=3.6$. The blue solid line shows the total interloper power spectrum from the 7 CO interloper lines. The 7 dashed lines show the interloper power spectrum for the 7 CO interloper lines respectively.}
    \label{fig:sphericalpower}
\end{figure}

\begin{figure}
	\includegraphics[width=\columnwidth]{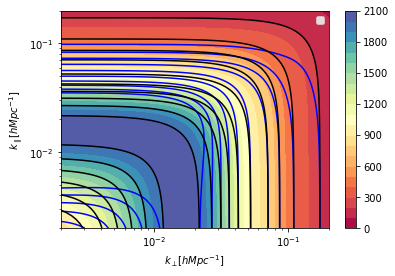}
    \caption{Contours of constant power in the $k_\perp-k_\parallel$ plane for
 the target $[\rm CII]$ power spectrum at $z_t=3.6$. The black contours neglect redshift space distortions, while the black contours and color-scale include them.The colorbar is in units of $\mu K^2 (\rm Mpc/h)^3$ }
    \label{fig:ciian}

\includegraphics[width=\columnwidth]{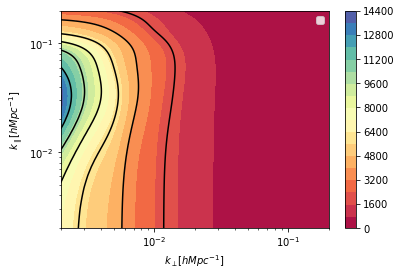}
    \caption{The contour with lowest power is at $P(k) = 2 \times 10^3 \mu K^2 (\rm Mpc/h)^3$ and the contours increase inwards as $\Delta P = 2 \times 10^3 \mu K^2 (\rm Mpc/h)^3$. The black contours illustrate the anisotropy of the summation of the seven interloper line power spectrum including redshift space distortion and coordinate mapping distortion. The $k_\parallel$ direction is strongly elongated}
    \label{fig:interloperan}
\end{figure}

 \begin{table}
    \centering
    \caption{Redshifts $z_j$, intensities $\langle I\rangle$ (in units of $\mu K$) and line biases $b_{\rm line}$ for the 7 interloper \rm CO lines}
    \label{tab:values}
    \begin{tabular}{|c|c|c|c|c|c|c|c|c|} 
\hline 
 J& 4 &5 &6 &7 & 8 & 9 & 10\\ 
 $z_j$& 0.113 &0.391 &0.670 &0.948 &1.226 &1.504 &1.783\\
 $\langle I\rangle$  & 0.07 &
 0.08 & 0.10 & 0.06 & 0.04 & 0.03 & 0.02
\\ $b_{\rm line}$& 0.6 &0.8 &1 &1.2 &1.3 &1.4 &1.6\\
\hline
\end{tabular}
\end{table}According to the values in Table~\ref{tab:values}, we plot spherically averaged $k^3 P(k)/(2\pi)^2$ of the $[\rm CII]$ line at $z=3.6$ and the 7 CO lines that interlope with it in Figure~\ref{fig:sphericalpower}.
As in previous work \citep{2016ApJ...825..143L, 2012ApJ...745...49G}, we plot contours of constant power in the $k_\perp-k_\parallel$ plane in Figure~\ref{fig:ciian} and~\ref{fig:interloperan} to visualize the anisotropy of the sum of the seven interloper power spectrum. Figure~\ref{fig:ciian} characterizes the redshift space distortion in the target $[\rm CII]$ line power spectrum. The contours in Figure~\ref{fig:interloperan} illustrate the anisotropy of the power spectrum summation of the interloper CO lines from the coordinate mapping distortion, with the $k_\parallel$ strongly elongated.
\section{PARAMETER ESTIMATION FORMALISM}\label{S:PEF}
The first part of this section describes the Fisher matrix forecast methodology. In the second part we derive the interloper bias to the measured power spectrum. 
\subsection{Fisher matrix forecast}
We use the Fisher matrix methodology to forecast how well a hypothetical $[\rm CII]$ survey will constrain $f_{NL}$. We first consider the simple case with only 1 CO interloper $\rm CO\left( 4 \rightarrow 3 \right)$. We consider a parameter space \textbf{q}. Our forecast is conducted at $z_t = 3.6$.\par
We investigate whether the $f_{NL}$ and $[\rm CII]$ target line, which is characterized by $I_{\rm CII}$ and $b_{\rm CII}$ can be forecasted precisely with the interloper contamination which is characterized by $\langle I_{\rm CO\left( 4 \rightarrow 3 \right)}\rangle$ and $b_{\rm CO\left( 4 \rightarrow 3 \right)}$ by calculating the Fisher matrix with component for the $i\rm th$ and $j \rm th$ of
$\textbf{q}$\\\begin{eqnarray}
\label{eq:fisher}F_{ij}&=&\int_{-1}^1{\rm d}\mu\int_{k_{\rm min}}^ {k_{\rm max}}  \frac{ k^2 {\rm d}k }{8 \pi^2} \ V_{\rm eff}(k,\mu,z_t) \frac{\partial {\rm ln} P_{\rm tot}(k,\mu,z_t)}{\partial q_{i} }\nonumber\\
&&\times\frac{\partial {\rm ln} P_{\rm tot}(k,\mu,z_t)}{\partial q_{j} }
\end{eqnarray}
where the total power spectrum including the $[\rm CII]$ and $\rm CO\left( 4 \rightarrow 3 \right)$ power is given as Eq.~(\ref{eq:ptot}). $V_{\rm eff}$ is the effective volume of the redshift bin at $z_t = 3.6$, $V_{\rm eff} =   \left[\frac{P_{\rm tot}(k,\mu,z_t)}{P_{\rm tot}(k,\mu,z_t) + P_{\rm shot}(z_t) + \tilde P_N(k,\mu)} \right]^2 V_i$, $P_{\rm shot}(z_t)$ is given as Eq.~(\ref{eq:pshot}), $\tilde P_N(k,\mu)$ is the effective instrumental noise power spectrum given later as Eq.~(\ref{eq:effpn}), $V_i$ is the volume of the redshift bin between $z_{min}$ and $z_{max}$ with a fraction of the sky $f_{sky}$; specifically, $V_i=\dfrac{4\pi}{3}f_{sky}[\chi^3(z_{max})-\chi^3(z_{min})]$, where $\chi(z)$ is the co-moving distance to redshift z.
\subsection{Parameter bias forecast}
The bias from interloper contamination in power spectrum detection can be calculated using the formalism constructed in \citet{2016PASJ...68...12P}. 
The parameter bias is 
\begin{equation}
    \label{eq:parabias}
    \Delta\textbf{q} = \textbf{F}^{-1} \Delta\textbf{D}
\end{equation}
where F is the Fisher matrix with the parameter space of \textbf{q}. The $j$th component of $\Delta \textbf{D}$ is given as 
\begin{eqnarray}
    \Delta D_j &=& \int_{-1}^1{\rm d}\mu\int_{k_{\rm min}}^ {k_{\rm max}}  \frac{ k^2 {\rm d}k }{8 \pi^2} \ V_{\rm eff}(k,\mu,z_t) \frac{\Delta P_{t}(k,\mu,z_t)}{P_t}\nonumber\\
    &&\times\frac{\partial {\rm ln} P_t(k,\mu,z_t)}{\partial q_{j} }
\end{eqnarray}
where $\Delta P_{t}(k,\mu,z_t)$ is the bias the interlopers contribute to the target power spectrum $P_{t}(k,\mu,z_t)$, i.e. the second term in Eq.~(\ref{eq:pshot}).
\section{FIDUCIAL SURVEY AND RESULTS}\label{S:FSR}
Having quantified the PNG contribution to the $[\rm CII]$ line power spectrum and its interloper contamination, we describe our hypothetical fiducial survey in the first part of this section, and present forecasts for using this to remove the CO interloper contributions from the $[\rm CII]$ local shape PNG power spectrum detection in the second part. 
\subsection{Fiducial survey description}

As in \citet{2019ApJ...872..126M}, we consider an Planck-like telescope having an aperture with dish length of $D_{ant} = 1.5$ m.  We assume a frequency range of 310-620 GHz, corresponding to a [CII] redshift range of $z=2.06-5.13$, and a frequency resolution $\Delta\nu = 0.4$ GHz, aperture temperature $T_{\rm aperture} = 40$ K, and approximately 
$N_{\rm det} = 10^4$ detectors working for $\tau_{\rm tot} = 4\times 10^4$ hours. We assume a survey sky coverage fraction $f_{\rm sky} = 0.34$, which leads to an angular limit on the sky of $\theta_{max} = \sqrt{\Omega_{\rm surv}} = \sqrt{4\pi f_{\rm sky}} = 118^{\circ}$. 

We include in our noise model photon noise from the CMB ($T_{\rm B}=2.725$ K), galactic dust ($T_{\rm B}=18$ K) and  zodiacal dust ($T_{\rm B}=240$ K) contributions. The emissivity of the dust is $\varepsilon = \varepsilon_{\circ}(\nu/\nu_{\circ})^{\beta}$,
where for galactic dust emission $\varepsilon_{\circ}=2\times10^{-4}$, $\nu_{\circ}=3$ THz, $\beta=2$;  while for zodiacal dust emission $\varepsilon_{\circ}=3\times10^{-7}$, $\nu_{\circ}=2$ THz, $\beta=2$. The radiation emitted at frequency $\nu$ is
\begin{equation}
I_{\nu}=\varepsilon(\nu)B_{\nu}(T)=\varepsilon(\nu)\frac{2h\nu^3}{c^2}\frac{1}{e^{\frac{h\nu}{k_B T_B}}-1}\, .
\end{equation}
The Rayleigh-Jeans Law yields 
\begin{equation}
T_{dust}=\frac{\varepsilon(\nu)B_{\nu}(T)c^2}{2\nu^2k_B}=\varepsilon(\nu)h\nu\frac{1}{k_B(e^{\frac{h\nu}{k_B T_B}}-1)}
\label{eq:dust}
\end{equation}
From Eq.~(\ref{eq:dust}) we get that at the observed frequency for the $[\rm CII]$ line, $T_{\rm galactic\: \rm dust}=46\mu K$ and $T_{\rm zodiacal \: \rm dust}=53\mu K$; therefore the systematic temperature is
\begin{equation}
 T_{sys}=T_{\rm aperture}+T_{\rm CMB}+T_{\rm zodiacal \: \rm dust}+T_{\rm galactic \: \rm dust}=42.725K 
\end{equation}
The instrumental noise for each k-mode is \citep{2019ApJ...870L...4M}
\begin{equation}\label{eq:noise}
P_{\rm N} = \frac{T_{\rm sys}^2}{\tau_{\rm tot}N_{\rm det}}\Omega_{\rm surv}\left(\frac{dl}{d\theta}\right)^2 \frac{dl}{d\nu}.
\end{equation}

The scale-independent correction to the clustering bias is relatively negligible at small scales, so we set $k_{\rm max} = 0.1 h\:Mpc^{-1}$.\par

We consider the largest scales recoverable from foreground contamination.  We model this with the parameter $\eta_{\rm min}$, defined as the ratio of the observed frequency divided
by the maximum bandwidth over which the frequency dependent
response of the instrument is assumed to be smooth, as in \citet{2019ApJ...872..126M}. We model the recoverable largest-scale per-parallel mode from foregrounds as $k_{\parallel,\rm min} = 2\pi\eta_{\rm min}[\nu_{obs}dl/dv]^{-1}$. Setting $\eta_{\rm min} = 4.5$, which is 3$\times$ the frequency-bandwidth ratio of the fiducial instrument, we have $k_{\parallel,\rm min} = 0.0077$ Mpc$^{-1}$. The survey area sets the largest-scale perpendicular mode, given by $k_{\bot,\rm min} = 2\pi[2\sin(\theta_{\rm max}/2)\frac{dl}{d\theta}]^{-1} = 0.0005$ Mpc$^{-1}$. The smallest-scale modes are generated by the spatial and spectral resolutions respectively, $k_{\bot,max}\approx 2\pi[\dfrac{c/\nu_{obs}}{D_{ant}}\frac{dl}{d\theta}]^{-1}$, $k_{\parallel,\rm max}\approx 2\pi[\delta\nu\frac{dl}{dv}]^{-1}$.

We introduce the attenuation of the signal due to foregrounds as in \citet{2019ApJ...872..126M}
\begin{equation}
\gamma_{\rm min}(k_{\perp},k_{\parallel}) = \left (1-e^{-k_{\perp}^{2}/(k_{\perp,{\rm min}}/2)^{2}}\right)  
\times \left (1-e^{-k_{\parallel}^{2}/(k_{\parallel,{\rm min}}/2)^{2}} \right)\, ,
\end{equation}
and due to finite spectral and angular resolution
\begin{equation}
\gamma_{\rm max}(k_{\perp},k_{\parallel}) = e^{-(k_{\perp}{^2}/k_{\perp,{\rm max}}^{2}+k_{\parallel}{^2}/k_{\parallel,{\rm max}}^{2})}\, ,
\end{equation}
writing the effective instrumental noise as
\begin{equation}
\label{eq:effpn}
\tilde P_{\rm N}(k,\mu,z) = P_{\rm N} \gamma^{-1} _{\rm max}(k,\mu)\gamma^{-1} _{\rm min}(k,\mu)\, .
\end{equation}

\subsection{Forecast results}
In this section we forecast the local shape $f_{NL}$ measured from only the auto-power spectrum. The Fisher analysis results for this section are listed in Table~\ref{tab:loc}.\par \begin{table}
    \centering
    \caption{Fisher analysis results for $f_{NL}^{loc}$}
    \label{tab:loc}
    \begin{tabular}{|c|c|c|c|} 
\hline 
 Interloper number& 0 &1 &7\\ 
 $\sigma(f_{NL}^{loc})$ for auto spectrum& 1.32 &1.64 &4.04\\
 $\sigma(f_{NL}^{loc})$ for cross-correlation  & 0.65 &
 0.69 & 0.85\\
\hline
\end{tabular}
\end{table}

\begin{figure}
\includegraphics[width=\columnwidth]{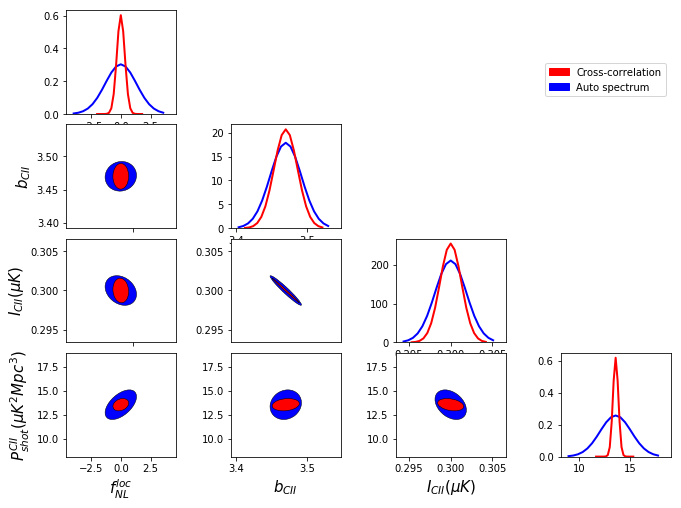}
    \caption{$1-\sigma$ forecast of the constraint on $f_{NL}^{loc}$ for the Fisher analysis without considering interlopers.}
    \label{fig:loc1}
\end{figure}

In the first scenario we assume there is no interloper contamination. The parameter space includes four parameters $\textbf{q}=\{f_{NL}^{loc}, \langle I_{\rm CII}\rangle, b_{\rm CII},P^{\rm CII}_{\rm shot}\}$. We set the fiducial value of the parameters as $f_{NL}^{loc} = 0$, $\langle I_{\rm CII}\rangle = 0.30 \mu K$, $b_{\rm CII} = 3.47$, as what was calculated before, $P^{\rm CII}_{\rm shot} = 13.57 \mu K^2 (\rm Mpc/h)^3$. We find $\sigma(f_{NL}^{loc}) = 1.32$, which being of order $\mathcal{O}(1)$ is the target level of uncertainty needed to discriminate between inflation models. This result deviates from the one from \citet{2019ApJ...872..126M} because we adopted different fiducial survey parameters, and therefore yield a smaller instrumental noise. We plot the $1-\sigma$ errors in different parameter planes shown as the yellow ellipses in Figure~(\ref{fig:loc1}).\par

\begin{figure*}
\includegraphics[width=0.9\textwidth]{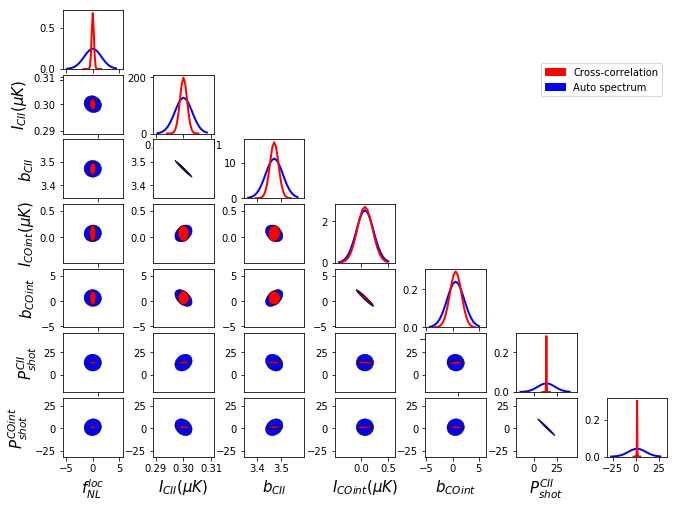}
    \caption{$1-\sigma$ forecast of the constraint on $f_{NL}^{loc}$ for the Fisher analysis with considering $\rm CO\left( 4 \rightarrow 3 \right)$ interloper.}   \label{fig:loc2}
\end{figure*}

In the second scenario we include a contamination from one interloper $\rm CO\left( 4 \rightarrow 3 \right)$ at $z_j = 0.113$ according to the second term in Eq.~(\ref{eq:ptot}), the anisotropic contribution to the [CII] power spectrum due to $\rm CO\left( 4 \rightarrow 3 \right)$. The parameter space now includes seven parameters $\textbf{q}=\{f_{NL}^{loc}, \langle I_{\rm CII}\rangle, b_{\rm CII}, \langle I_{\rm CO\left( 4 \rightarrow 3 \right)}(z_j)\rangle, b_{\rm CO\left( 4 \rightarrow 3 \right)}(z_j),P^{\rm CII}_{\rm shot},$ $P^{\rm CO\left( 4 \rightarrow 3 \right)}_{\rm shot}(z_j)\}$. We set the fiducial values of the interloper parameters as $\langle I_{\rm CO\left( 4 \rightarrow 3 \right)\rangle}(z_j)\rangle = 0.07 \mu K$, $b_{\rm CO\left( 4 \rightarrow 3 \right)}(z_j) = 0.6$ as listed in Table~\ref{tab:values}, $P^{\rm CO\left( 4 \rightarrow 3 \right)}_{\rm shot}(z_j) = 1.35 \mu K^2 (\rm Mpc/h)^3$, and the remaining parameters are set as in the first scenario. We compute the marginalized $\sigma(f_{NL}^{loc}) = 1.64$, a 24\% increase due to interloper contamination.  We plot the $1-\sigma$ errors in different parameter planes shown as yellow ellipses in Figure~(\ref{fig:loc2}). Alternatively, if we consider the interloper contamination as pure noise other than mixing signals for the target $\rm CII$ line, the parameter space becomes $\textbf{q}=\{f_{NL}^{loc}, \langle I_{\rm CII}\rangle, b_{\rm CII}, P^{\rm CII}_{\rm shot}\}$ and we find $\sigma(f_{NL}^{loc}) = 1.60$. Note that treating the CO(4-3) line as noise gives you slightly less error in $f_{NL}^{loc}$ because this model has less degrees of freedom.  The uncertainty in $f_{NL}^{loc}$ becomes much larger when multiple interloper lines are introduced. If we include interloper contamination from all the 7 CO interlopers whose fiducial intensities and biases are given in Table~\ref{tab:values} and marginalize over all these fiducial quantities, we get $\sigma(f_{NL}^{loc}) = 4.04$.\par
Considering the case where one interloper $P_{\rm CO\left( 4 \rightarrow 3 \right)}$ biases the measured power spectrum, we get the parameter bias $\Delta \textbf{p}^T = [\Delta f_{NL}^{loc}, \Delta b_{\rm CII}, \Delta \langle I_{\rm CII}\rangle,\Delta  P_{\rm shot}^{\rm CII}] = [4.54,-5.45,0.49,6.87]$.  This would suggest one interloper can masquerade as a 2.8$\sigma$ detection of $f_{NL}$ while providing a false positive for non-standard inflation models.  This result shows that the presence of interlopers are prohibitive when attempting to measure $f_{NL}^{loc}$ using an auto-power spectrum.
\section{CROSS-CORRELATION WITH OTHER LINES}\label{S:CROSS}
According to our Fisher analysis results, interloper contamination highly biases local PNG constraints. Next, we consider cross-correlating the target $[\rm CII]$ line with a data cube centered on a different frequency corresponding to a target redshift $z_t = 3.6$ for the $\rm CO\left( 4 \rightarrow 3 \right)$ line, which could reduce the interloper contamination to the PNG constraint. We label the auto power spectra of the $[\rm CII]$ and $\rm CO\left( 4 \rightarrow 3 \right)$ at $z_t = 3.6$ as $P_{\rm CII}(k,\mu)$ and $P_{\rm CO\left( 4 \rightarrow 3 \right)}(k,\mu)$, respectively. The cross-power spectrum between the $[\rm CII]$ and $\rm CO\left( 4 \rightarrow 3 \right)$ data cubes is
\begin{align}
P_{\rm x}(k, \mu) =& I_{\rm CII} I_{\rm CO\left( 4 \rightarrow 3 \right)} b_{\rm CII} b_{\rm CO\left( 4 \rightarrow 3 \right)}\left(1 + \beta_{\rm CII} \mu^2 \right)\nonumber \\
&\times \left(1 + \beta_{\rm CO\left( 4 \rightarrow 3 \right)} \mu^2 \right){\rm exp}\left(-\frac{k^2 \mu^2 \sigma_v^2}{H^2(z)}\right)P_{0}(k,z_t) + P_{\rm shot}^X
\label{eq:pcross}
\end{align}
The shot noise in the cross-power spectrum is
\begin{align}
P_{\rm shot}^X(z) = \frac{c^4 f_{\rm duty}}{4k_B^2 \nu_{obs \rm CII}^2 \nu_{obs \rm CO\left( 4 \rightarrow 3 \right)}^2}  \int_{M_{\rm min}}^{M_{\rm max}} dM \frac{dn}{dM}\nonumber \\ \times {\left[\frac{L(M,z)}{4 \pi \mathcal D_L^2} 
\left ( \frac{dl}{d\theta} \right )^{2} \frac{dl}{d\nu} \right ]}_{\rm CII}{\left[\frac{L(M,z)}{4 \pi \mathcal D_L^2} 
\left ( \frac{dl}{d\theta} \right )^{2} \frac{dl}{d\nu} \right ]}_{\rm CO\left( 4 \rightarrow 3 \right)}.
\end{align}
Let the observables be the data vector \textbf{d} = $\{P_{\rm CII}(k,\mu), P_{\rm CO\left( 4 \rightarrow 3 \right)}(k,\mu), P_x \}$.  We consider a parameter space \textbf{q} at $z_t = 3.6$. The Fisher matrix element with component for the $i\rm th$ and $j \rm th$ of $\textbf{q}$ is \citep{Tegmark:1996bz}   
\begin{equation}
\label{eq:fisher2}
F_{ij}=V_i\int_{-1}^1{\rm d}\mu \int_{k_{\rm min}}^ {k_{\rm max}}  \frac{ k^2 {\rm d}k}{8 \pi^2} \frac{\partial \textbf{d}}{\partial q_{i} } \boldsymbol{\Xi}^{-1} \frac{\partial \textbf{d}^T}{\partial q_{j} }
\end{equation}
 where the entries of the covariance matrix $\Xi$ are given in \citet{2009ApJ...690..252L}.  We can also use Eq.~\ref{eq:parabias} to compute the shift in the parameters due to interloper contamination.  In this case, the formula for $\Delta D$ becomes
\begin{equation}
\Delta D_j=V_i\int_{-1}^1{\rm d}\mu \int_{k_{\rm min}}^ {k_{\rm max}}  \frac{ k^2 {\rm d}k}{8 \pi^2} \Delta\textbf{d} \boldsymbol{\Xi}^{-1} \frac{\partial \textbf{d}^T}{\partial q_{j} }\, ,
\end{equation}
where $\Delta\textbf{d}$ is the shift in the data vector due to interloper contamination.

The Fisher analysis results for this section are listed in Table~\ref{tab:loc}.
 
 We first consider the scenario without interloper lines, i.e. only the first term in Eq.~\ref{eq:ptot} is considered. The parameter space is $\textbf{q}=\{f_{NL}^{loc}, \langle I_{\rm CII}(z_t)\rangle, b_{\rm CII}(z_t),$ $ P^{\rm CII}_{\rm shot}(z_t), \langle I_{\rm CO\left( 4 \rightarrow 3 \right)}(z_t)\rangle, b_{\rm CO\left( 4 \rightarrow 3 \right)}(z_t), P^{\rm CO\left( 4 \rightarrow 3 \right)}_{\rm shot}(z_t),P^X_{\rm shot}\}$, in which the fiducial values for the $f_{NL}^{loc}$, $\langle I_{\rm CII}(z_t)\rangle$, $b_{\rm CII}(z_t)$ are set as in in Section \ref{S:FSR}, $\langle I_{\rm CO\left( 4 \rightarrow 3 \right)}(z_t)\rangle$ and $b_{\rm CO\left( 4 \rightarrow 3 \right)}(z_t)$ are calculated from the method discussed in Section \ref{S:PNGLIM}, $\langle I_{\rm CO\left( 4 \rightarrow 3 \right)}(z_t)\rangle = 0.60 \mu K$, $b_{\rm CO\left( 4 \rightarrow 3 \right)}(z_t) = 2.8$, $P^{\rm CO\left( 4 \rightarrow 3 \right)}_{\rm shot}(z_t)= 88.01 \mu K^2 (\rm Mpc/h)^3$, $P^{X}_{\rm shot}= 32.13 \mu K^2 (\rm Mpc/h)^3$. We use the $8\times8$ Fisher matrix from Eq.~\ref{eq:fisher2} to find the constraint on the parameters. $1-\sigma$ errors in different parameter planes are shown as the yellow ellipses in Figure~\ref{fig:loc1}. In this Fisher analysis assuming the dataset $\{P_{\rm CII}(k,\mu), P_{\rm CO\left( 4 \rightarrow 3 \right)}(k,\mu), P_x \}$, we find $\sigma(f_{NL}^{loc}) = 0.65$, much lower than the initial experimental setup value for the $[\rm CII]$ auto power spectrum of $\sigma(f_{NL}^{loc}) = 1.32$.  This result is expected since having multiple lines add more constraining power, and we benefit from the multi-tracer cosmic variance cancellation, as addressed in \citet{2021PhRvD.103f3520L}.  \par

In the second scenario we include the contamination from one interloper $\rm CO\left( 4 \rightarrow 3 \right)$ at redshift 0.113 according to the second term in Eq.~\ref{eq:ptot}, the anisotropic contribution to the [CII] power spectrum due to $\rm CO\left( 4 \rightarrow 3 \right)$. 

The parameter space includes eleven parameters $\textbf{q}=\{f_{NL}^{loc}, \langle I_{\rm CII}(z_t)\rangle, b_{\rm CII}(z_t), P^{\rm CII}_{\rm shot}(z_t), \langle I_{\rm CO\left( 4 \rightarrow 3 \right)}(z_t)\rangle,$ $b_{\rm CO\left( 4 \rightarrow 3 \right)}(z_t), P^{\rm CO\left( 4 \rightarrow 3 \right)}_{\rm shot}(z_t), \langle I_{\rm int\:CO\left( 4 \rightarrow 3 \right)}(z_j)\rangle, b_{\rm int\: CO\left( 4 \rightarrow 3 \right)}(z_j),$ $ P^{\rm CO\left( 4 \rightarrow 3 \right)}_{\rm shot}(z_j)
, P^{X}_{\rm shot}
 \}$, in which the fiducial values $\langle I_{\rm int\:CO\left( 4 \rightarrow 3 \right)}(z_j)\rangle$ and $\:b_{\rm int\: CO\left( 4 \rightarrow 3 \right)}(z_j)$ for the interloper parameters are set as section 4.2. We use the $11 \times11$ Fisher matrix from Eq.~\ref{eq:fisher2} to find the constraint on the parameters. $1-\sigma$ errors in different parameter planes are shown as the red ellipses in Figure~\ref{fig:loc2}. We find that in this Fisher analysis $\sigma(f_{NL}^{loc}) = 0.69$, which is only 6\% more than the predicted $f_{NL}$ error in the case with no interlopers.  Note this increase is 4x less than the increase in the $f_{NL}$ error from the auto-power spectrum constraint.   We also see that the error for the two cross-correlation cases, one with no interlopers and one with interlopers, are equal within less than 6\%, suggesting that interlopers do not significantly bias measurements of $f_{\rm NL}$ when the auto and cross-power spectra are used jointly. Alternatively, if we consider the interloper contamination as pure noise other than mixing signals for the target $\rm CII$ line, we find $\sigma(f_{NL}^{loc}) = 0.78$, the parameter bias $\Delta \textbf{p}^T = \Delta[f_{NL}^{loc}, \langle I_{\rm CII}(z_t)\rangle, b_{\rm CII}(z_t), \langle I_{\rm CO\left( 4 \rightarrow 3 \right)}(z_t)\rangle, b_{\rm CO\left( 4 \rightarrow 3 \right)}(z_t),$ $P_{\rm shot}^{\rm CII}(z_t),P_{\rm shot}^{\rm CO(4 \rightarrow 3)}(z_t)] = [3.29, 0.21, -2.29, -0.037, 0.15, 6.77,$ $30.89]$. This results shows that including the interlopers as pure noise other than signals can not help to remove the interloper contamination. If we assume the dataset only includes the cross-spectrum $\{P_x \}$, we find $\sigma(f_{NL}^{loc}) = 0.85$, slightly higher than that from doing the cross-correlation considering auto-spectrum. This result shows that including the auto-spectrum does not aid much when removing the interlopers if you treat it as noise. \par
If we include all the 7 interlopers $\rm CO\left( 4 \rightarrow 3 \right)$ to $\rm CO\left( 10 \rightarrow 9 \right)$ at lower redshift, with all their intensities and biases being added to the parameter space \textbf{q}, in the cross-correlated Fisher analysis we get $\sigma(f_{NL}^{loc}) = 0.85$, dropping significantly as compared to $\sigma(f_{NL}^{loc}) = 4.04$ for the $[\rm CII]$ auto power spectrum Fisher analysis.

\section{Constraints ON Orthogonal Shape PNG}\label{S:Orth}
In this section we derive how interlopers affect constraints on measurements of orthogonal shape PNG.  We present the forecasts for the removal of CO interloper contributions to the $[\rm CII]$ auto power spectrum detection in the first subsection. In the second subsection we present the Fisher forecasts for the cross-correlation between $[\rm CII]$ line and other lines at the same redshift. The Fisher analysis results for this section are listed in Table~\ref{tab:orth}.\par \begin{table}
    \centering
    \caption{Fisher analysis results for $f_{NL}^{orth}$}
    \label{tab:orth}
    \begin{tabular}{|c|c|c|} 
\hline 
 Interloper number& 0 &1 \\ 
 $\sigma(f_{NL}^{orth})$ for auto spectrum& 52.75 &7.15 \\
 $\sigma(f_{NL}^{orth})$ for cross-correlation  & 42.24 &
 7.05 \\
\hline
\end{tabular}
\end{table}

\subsection{Auto power spectrum Fisher forecast}
We forecast the orthogonal shape PNG rather than the equilateral PNG since the former brings larger correction to the clustering bias, and thus a stronger constraint on $f_{NL}$. We set the fiducial value $f_{NL}^{orth}=0$. We adopt the same fiducial survey and remaining parameter values as described in Section \ref{S:FSR}. 

\begin{figure}
\includegraphics[width=\columnwidth]{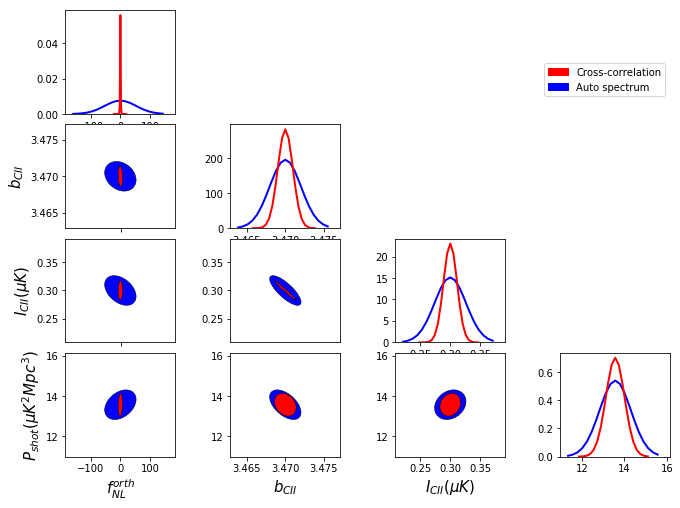}
    \caption{$1-\sigma$ forecast of the constraint on $f_{NL}^{orth}$ for the Fisher analysis without considering interlopers.}
    \label{fig:orth1}
\end{figure}

In the first scenario we assume there is no interloper contamination. The parameter space includes three parameters $\textbf{q}=\{f_{NL}^{orth}, \langle I_{\rm CII}\rangle, b_{\rm CII}, P^{\rm CII}_{\rm shot} \}$. We find $\sigma(f_{NL}^{orth}) = 52.75$, and we plot the $1-\sigma$ errors in different parameter planes shown as yellow ellipses in Figure~\ref{fig:orth1}.\par

In the second scenario we include the interloper contamination from one interloper $\rm CO\left( 4 \rightarrow 3 \right)$ at $z_j = 0.113$. The parameter space contains five parameters $\textbf{q}=\{f_{NL}^{orth}, \langle I_{\rm CII}(z_t)\rangle, b_{\rm CII}(z_t), P^{\rm CII}_{\rm shot}(z_t),  \langle I_{\rm CO\left( 4 \rightarrow 3 \right)}(z_j)\rangle, $ $b_{\rm CO\left( 4 \rightarrow 3 \right)}(z_j), P^{\rm CO\left( 4 \rightarrow 3 \right)}_{\rm shot}(z_j)\}$. We plot the $1-\sigma$ errors in different parameter planes shown as yellow ellipses in Figure~\ref{fig:orth2}. We find the marginalized $\sigma(f_{NL}^{orth}) = 42.24$. The reason for this error being smaller than that in the first scenario, as contrary to the case for $f_{NL}^{loc}$ is that $\Delta b_h^{orth}$ in ~(\ref{eq:biascorrection3}) are of the same sign for the target line and the interloper, while $\Delta b_{h}^{loc}$ in ~(\ref{eq:biascorrection2}) are of different signs.\par
Considering the case where one interloper $P_{\rm CO\left( 4 \rightarrow 3 \right)}$ biases the measured power spectrum, we compute the parameter bias $\Delta \textbf{p}^T = [\Delta f_{NL}^{orth}, \Delta b_{\rm CII}, \Delta \langle I_{\rm CII}\rangle,\Delta P_{\rm shot}^{\rm CII}] = [49.20, -5.58,0.51,6.25]$.  This points to a modest bias relative to the predicted errors, in that the interlopers could masquerade as a 1.2-sigma measurement of $f_{NL}$.

\subsection{Cross-correlation Fisher forecast}
We adopt the same fiducial survey and value of parameters for cross-correlation as in Section \ref{S:CROSS}. 
We first consider the scenario without interloper lines with the parameter space being $\textbf{q}=\{f_{NL}^{orth}, \langle I_{\rm CII}(z_t)\rangle, b_{\rm CII}(z_t), P^{\rm CII}_{\rm shot}(z_t), \langle I_{\rm CO\left( 4 \rightarrow 3 \right)}(z_t)\rangle,$ $ b_{\rm CO\left( 4 \rightarrow 3 \right)}(z_t), P^{\rm CO\left( 4 \rightarrow 3 \right)}_{\rm shot}(z_t)
, P^{X}_{\rm shot}
\}$. $1-\sigma$ errors in different parameter planes are shown as red ellipses in Figure~\ref{fig:orth1}. We find that in this cross-correlated Fisher analysis $\sigma(f_{NL}^{orth}) = 7.05$, dropping significantly as compared to $\sigma(f_{NL}^{orth}) = 52.75$ for the $[\rm CII]$ auto power spectrum Fisher analysis. \par
\begin{figure*}
\includegraphics[width=0.9\textwidth]{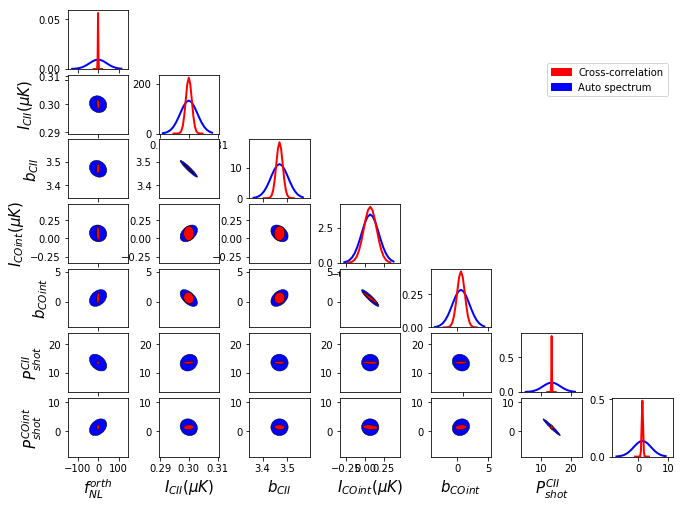}
    \caption{$1-\sigma$ forecast of the constraint on $f_{NL}^{orth}$ for the Fisher analysis with considering $\rm CO\left( 4 \rightarrow 3 \right)$ interloper.}
    \label{fig:orth2}
\end{figure*}

In the second scenario we include the interloper contamination from one interloper $\rm CO\left( 4 \rightarrow 3 \right)$ at redshift 0.113. The parameter space includes seven parameters $\textbf{q}=\{f_{NL}^{orth}, \langle I_{\rm CII}(z_t)\rangle, b_{\rm CII}(z_t), P^{\rm CII}_{\rm shot}(z_t), \langle I_{\rm CO\left( 4 \rightarrow 3 \right)}(z_t)\rangle, $ $b_{\rm CO\left( 4 \rightarrow 3 \right)}(z_t), P^{\rm CO\left( 4 \rightarrow 3 \right)}_{\rm shot}(z_t),\langle I_{\rm int\:CO\left( 4 \rightarrow 3 \right)}(z_j)\rangle, b_{\rm int\: CO\left( 4 \rightarrow 3 \right)}(z_j), $ $P^{\rm CO\left( 4 \rightarrow 3 \right)}_{\rm shot}(z_j),P^{X}_{\rm shot} \}$. $1-\sigma$ errors in different parameter planes are shown as red ellipses in Figure~\ref{fig:orth2}. We note that in this cross-correlated Fisher analysis $\sigma(f_{NL}^{orth}) = 7.05$, dropping significantly as compared to $\sigma(f_{NL}^{orth}) = 42.24$ for the $[\rm CII]$ auto power spectrum Fisher analysis. Similar to the local PNG case, cross-power spectra tend to remove the effect of interlopers when measuring $f_{\rm NL}$. Alternatively, if we consider the interloper contamination as pure noise other than mixing signals for the target $\rm CII$ line, we find $\sigma(f_{NL}^{orth}) = 7.29$, the parameter bias $\Delta \textbf{p}^T = [f_{NL}^{orth}, \langle I_{\rm CII}(z_t)\rangle, b_{\rm CII}(z_t), \langle I_{\rm CO\left( 4 \rightarrow 3 \right)}(z_t)\rangle, b_{\rm CO\left( 4 \rightarrow 3 \right)}(z_t), $ $P_{\rm shot}^{\rm CII}(z_t), P_{\rm shot}^{\rm CO}(z_t)] = [59.31, 0.47,-5.53, 0.041,-0.20,6.56,$ $30.84]$. This results shows that including the interlopers as pure noise other than signals can not help to remove the interloper contamination. If we assume the dataset only includes the cross-spectrum $\{P_x \}$, we find $\sigma(f_{NL}^{orth}) = 13.51$, larger than that from considering both the cross-spedtrum and the auto-spectra. This result shows that including the auto-spectrum is necessary for removing the interlopers when treating it as noise. \par

\section{DISCUSSION}\label{S:DISCUSSION}
In this section we discuss how the PNG constraints depend on astrophysical models.Specifically, we consider how both different modelling of CO line luminosity and the minimum mass of halos that emit lines affect the PNG constraints.\par
We set $M_{\rm min} = 10^{9}M_\odot$ in our base model.In the fiducial [CII] model we use throughout the paper, we set the minimum halo mass for [CII] line emission to be $M_{\rm min} = 10^{9}M_\odot$.  This parameter is very uncertain, with reasonable values between $M_{\rm min} = 10^{9-11}M_\odot$.  We find when setting $M_{\rm min} = 10^{10}M_\odot$ without considering interloper contamination, $\langle I_{\rm CII}(z_t)\rangle$ gets lower and therefore yields a weaker constraint. Specifically, for the case with only the auto-power spectrum only and no interlopers present, $\sigma(f_{NL}^{loc}) = 1.54$, a 17\% increase. For the case with both auto-power and cross-power spectrum, $\sigma(f_{NL}^{loc}) = 0.78$, a $13\%$ increase.\par
 For the fiducial CO model in this work we use the \citet{2016ApJ...817..169L} model for the interlopers. Including only one CO interloper $CO\left( 4 \rightarrow 3 \right)$ line contamination, we consider two other models with the CO luminosity linearly related to the halo mass. In model A considering $f_{\rm duty} = t_s/t_{\rm age}(z)$ in \citet{2013ApJ...768...15P}, we find a lower intensity for $CO\left( 4 \rightarrow 3 \right)$, and thus a stronger constraint on PNG. Specifically, for the auto-power only case with one interloper we find $\sigma(f_{NL}^{loc}) = 1.55$, a 6\% reduction; for the case with both auto-power and cross-power, $\sigma(f_{NL}^{loc}) = 0.67$, a $3\%$ reduction. In \citet{2016ApJ...830...34K}, we get higher intensity for $CO\left( 4 \rightarrow 3 \right)$, and thus a weaker constraint on PNG with $\sigma(f_{NL}^{loc}) = 1.76$ and $\sigma(f_{NL}^{loc}) = 0.83$ for cases with only auto-power and with both auto-power and cross-power spectrum respectively.\par
We use the luminosity model in  \citet{2010JCAP...11..016V} to get the intensity ratios between $CO\left( J \rightarrow J-1 \right)$ and $CO\left( 1 \rightarrow 0 \right)$. Considering instead the relationship $CO\left( J \rightarrow J-1 \right) = J^3 CO\left( 1 \rightarrow 0 \right)$ in \citet{2009ApJ...702.1321O} and get higher interloper intensities, and thus a weaker constraint on PNG with $\sigma(f_{NL}^{loc}) = 4.80$ if we include seven interlopers. While the relationship of $CO\left( J \rightarrow J-1 \right) = CO\left( 1 \rightarrow 0 \right)$ in \citet{2011ApJ...741...70L} yields lower interloper intensities, and thus a stronger constraint on PNG with $\sigma(f_{NL}^{loc}) = 1.34$ and $\sigma(f_{NL}^{loc}) = 0.60$ for cases with only auto-power and with both auto-power and cross-power spectrum respectively.

\section{CONCLUSION}\label{S:CONCLUSION}
Probing constraints on Primordial Non-Gaussianity (PNG) characterized by $f_{NL}$ is a strong discriminant among cosmological models. PNG leads to a scale-dependent clustering bias for emission lines. We consider $[\rm CII]$ line intensity mapping at the target redshift $z_t = 3.6$ to model how interlopers affect measurements of $f_{NL}^{loc}$ and $f_{NL}^{orth}$.\par
Interloper contamination from CO line confusion provides an important systematic concern for our intensity mapping method. We separate the anisotropic CO interloper contamination at the power spectrum level. We obtained constraints on $f_{NL}$ from a future survey with and without CO lines interloping, and found that the interloper contamination leads to weaker constraints on $f_{NL}$ along with significant bias to the parameter. \par
We model the cross-correlation between $[\rm CII]$ and CO(4-3) at the same redshift as a way to reduce the interloper contamination at target redshift. We model constraints on $f_{NL}$ by calculating auto power spectrum and cross power spectrum between the data cubes which contain $[\rm CII]$ and CO(4-3). We find the interloper contamination for the PNG probing can be largely removed using this method.\par
Primordial Non-Gaussianity will shed light on physics of primordial fluctuations. Line intensity mapping is a potential technique to probe PNG by providing complimentary information undetectable with other traditional galaxy surveys. Interloper contamination is an important systematic concern for this technique, as obtained by using the power anisotropy separation method; but it can be reduced largely by using cross-correlation techniques.

\section*{Acknowledgements}

We thank Patrick Breysse for his helpful feedback on an earlier version of this manuscript.  ARP was supported by the Simons Foundation and by NASA under award numbers 80NSSC18K1014 and NNH17ZDA001N.

\bibliographystyle{mnras}
\bibliography{example}

%%%%%%%%%%%%%%%%%%%%%%%%%%%%%%%%%%%%%%%%%%%%%%%%%%

%%%%%%%%%%%%%%%%% APPENDICES %%%%%%%%%%%%%%%%%%%%%

\appendix

%%%%%%%%%%%%%%%%%%%%%%%%%%%%%%%%%%%%%%%%%%%%%%%%%%

% Don't change these lines
\bsp	% typesetting comment
\label{lastpage}
\end{document}